\documentclass[multphys,vecphys]{svmult}

\usepackage{makeidx}         
\usepackage{graphicx}        
\usepackage{multicol}        
\usepackage[bottom]{footmisc}

\makeindex             

\newcommand{\ltsim}{\protect\raisebox{-0.5ex}{$\:\stackrel{\textstyle <}
        {\sim}\:$}}

\begin{document}

\title*{Physical parameters of evolved stars in clusters and in the field from line-depth ratios}
\titlerunning{Physical parameters in evolved stars}
\author{K. Biazzo\inst{1,2}\and L. Pasquini\inst{2}\and A. Frasca\inst{1}\and L. da Silva\inst{3} \and 
L. Girardi\inst{4}\and A. P. Hatzes\inst{5} \and J. Setiawan\inst{6} \and S. Catalano\inst{1} \and E. Marilli\inst{1}}
\authorrunning{K. Biazzo et al.}
\institute{Osservatorio Astrofisico di Catania-INAF, Catania, Italy
\texttt{kbiazzo@oact.inaf.it}
\and European Southern Observatory, Garching bei M\"unchen, Germany 
\and Observat\'orio Nacional-MCT, Rio de Janeiro, Brazil 
\and Osservatorio Astronomico di Padova-INAF, Padova, Italy 
\and Th\"uringer Landessternwarte, Tauterburg, Germany 
\and Max-Planck-Institute f\"ur Astronomie, Heidelberg, Germany 
}
%
%
\maketitle

\begin{abstract}
We present a high-resolution spectroscopic analysis of two samples of evolved stars selected in the 
field and in the intermediate-age open cluster IC~4651, for which detailed measurements of chemical 
composition were made in the last few years. Applying the Gray's method based 
on ratios of line depths, we determine the effective temperature and compare our 
results with previous ones obtained by means of the curves of growth of iron lines. 
The knowledge of the temperature enables us to estimate other 
fundamental stellar parameters, such as color excess, age, and mass.  
\end{abstract}

\section{Introduction}

The study of stellar populations in our Galaxy and in its neighborhoods has received in the last years 
a big impulse, especially thanks to the use of large telescopes and to the detailed spectroscopic analysis 
performed on high-resolution spectra. In this context, open clusters, that are homogeneous samples of stars 
having the same age and chemical composition, 
are very suitable to investigate the stellar and Galactic formation and evolution. 
In spite of this, the data on stars belonging to open clusters are often insufficient to adequately
constrain age, distance, metallicity, mass, color excess, and temperature.
This is due to the fact that the main classical tool to study cluster properties is the 
color-magnitude diagram, which suffers of several uncertainties and intrinsic biases due to, for example, 
the uncertain knowledge of the chemical composition and the reddening of the stars. As a consequence, spectroscopic 
methods, being independent of the reddening, are very efficient to evaluate temperatures of stars in clusters. 

Spectroscopic effective temperatures are usually determined imposing that the abundance of one chemical element 
with many lines in the spectrum (typically iron) does not depend on the excitation potential of the 
lines. Another method for determining effective temperature is based on line-depth ratios (LDRs).
It has been widely demonstrated that 
the ratio of the depths of two lines having different sensitivity to temperature is an excellent measure of stellar 
temperatures with a sensitivity as small as a few Kelvin degrees in the most favorable cases 
(\cite{Gray91,Gray01,Gray05}).

In the present paper, we apply the LDR method to high-resolution UVES and FEROS spectra for deriving effective 
temperatures in nearby evolved field stars with very good Hipparcos distances and in giants of the intermediate-age 
open cluster IC~4651. 
For both the star samples, the temperature was already derived spectroscopically, together with the
element abundances, with the curves of growth of absorption lines spread throughout the optical spectrum
(\cite{Pasqu04,daSil06}). In addition, for the stars belonging to the open cluster IC~4651, we make the first
robust determination of the average color excess, based on spectroscopic measurements.

\section{Star samples}
We have analysed seventy-one evolved field stars and six giant stars belonging to the open cluster 
IC~4651. The sample of field stars was already analysed by \cite{daSil06} for the determination of radii, 
temperatures, masses and chemical composition. The stars in the intermediate-age cluster IC~4651 have been 
selected from the sample studied by \cite{Pasqu04} for abundance estimates.

The field stars data were acquired with the FEROS 
spectrograph ($R=48\,000$) at the ESO 1.5m-telescope in La Silla (Chile), while the IC~4651 spectra were acquired 
with UVES ($R=100\,000$) at the ESO VLT Kueyen 8.2m-telescope in Cerro Paranal (Chile). In 
both cases, the signal-to-noise ratio ($S/N$) was  greater than 150 for all the spectra, which make 
them very suitable for the temperature determination described in Sec.~\ref{sec:Teff}.

\section{Effective temperature determination}
\label{sec:Teff}
The wavelength range covered by FEROS and UVES spectrographs contain a series of weak metal lines which can be used 
for temperature determination with the LDR method. Lines from similar elements such as iron, vanadium, titanium, but 
with different excitation potentials ($\chi$) have indeed different sensitivity to temperature. This is due to the 
fact that the line strength, depending on excitation and ionization processes, is a function of temperature and, to a 
lesser extent, of the electron pressure. For this reason, the better line couples are those with the largest 
$\chi$-difference. In the range 6150 \AA~$\ltsim \lambda \ltsim$ 6300\AA~there are several lines of this type 
whose ratios of their depths have been exploited for temperature calibrations (\cite{Gray91,Gray01,Cata02,Biazzo06b}), 
and for studies of the rotational modulation of the average effective temperature of magnetically active stars 
(\cite{Frasca05, Biazzo06a}) or for investigating the pulsational variations during the phases of a Cepheid star 
(\cite{Kovty00, Biazzo04}). In particular, we choose 15 line pairs for which \cite{Biazzo06b} already made suitable 
calibrations. 

\begin{figure*}
\centering
\vspace{-.5cm}
\includegraphics[width=5.2cm]{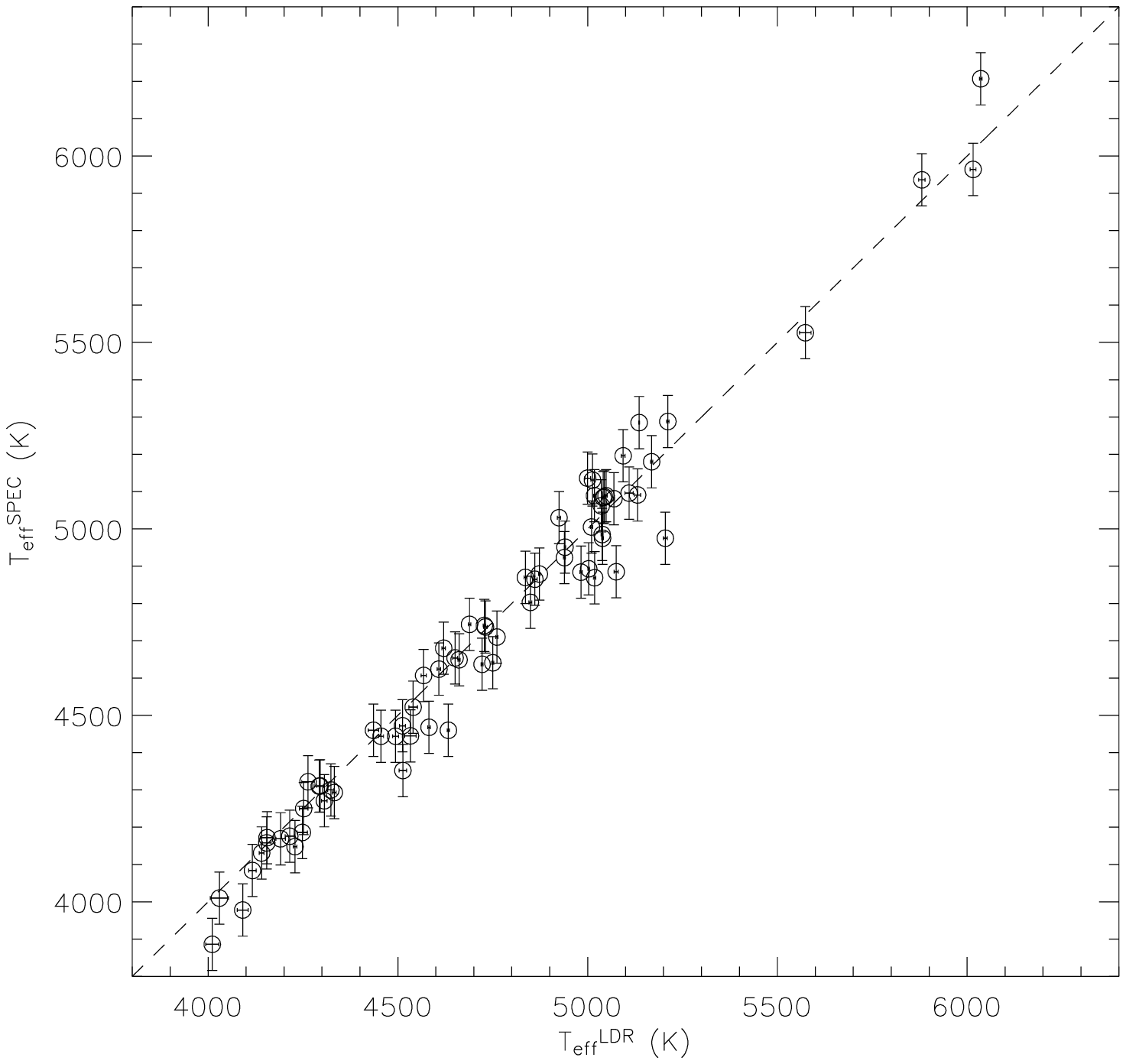}
\includegraphics[width=5.2cm]{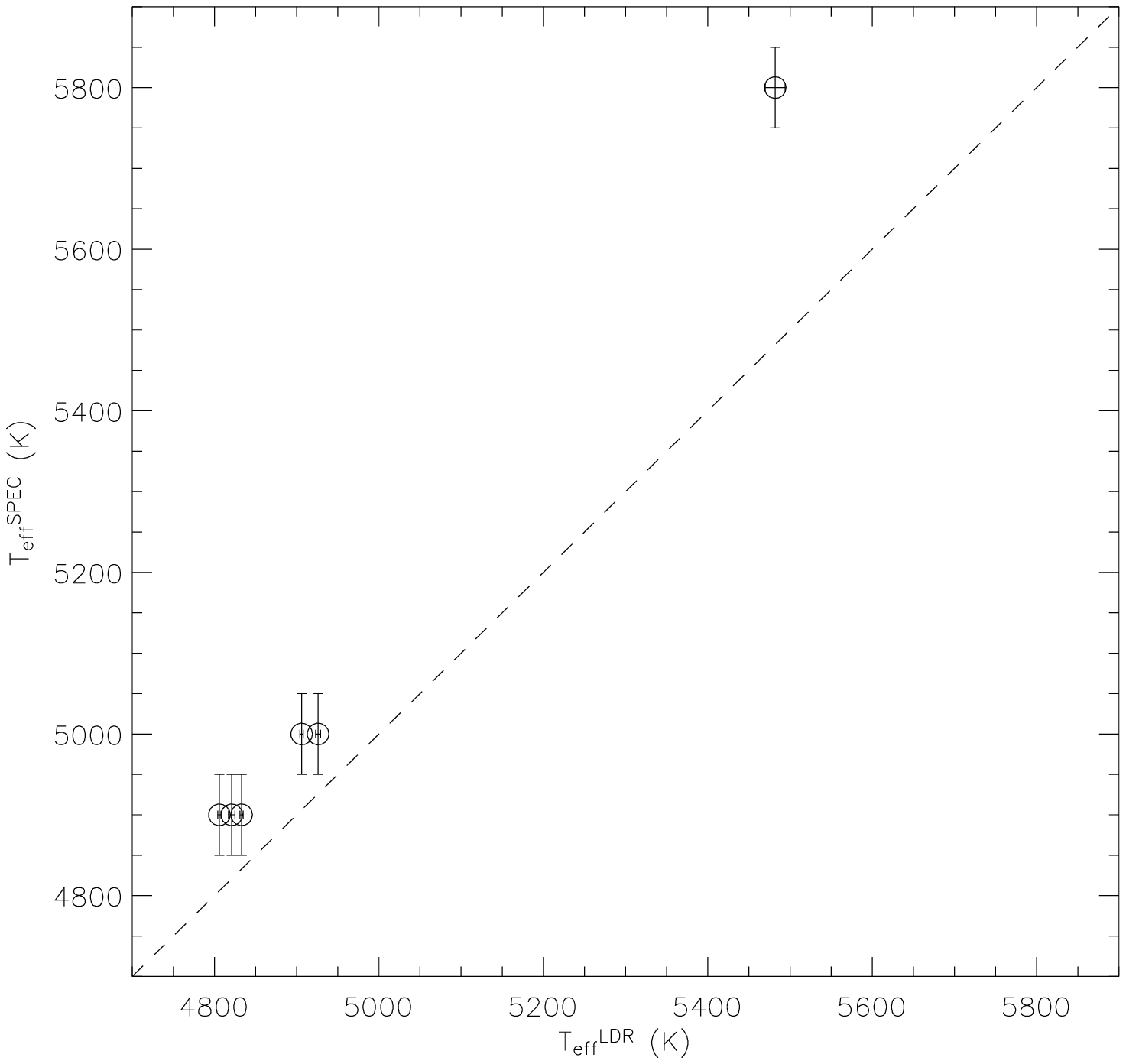} 
\caption{Comparison between the temperatures obtained from LDR and curve-of-growth 
analyses for the field stars ({\it left}) and the giants belonging to IC~4651 ({\it right}).}
\label{fig:evol_teff_comparison}
\end{figure*}

\paragraph{Field giant stars}
The comparison between the temperatures obtained by us ($T_{\rm eff}^{\rm LDR}$) and those obtained by \cite{daSil06} 
($T_{\rm eff}^{\rm SPEC}$) is plotted in Fig.~\ref{fig:evol_teff_comparison}. We find a very good agreement 
between $T_{\rm eff}^{\rm SPEC}$ and $T_{\rm eff}^{\rm LDR}$ in all the temperature range 4000--6000 K. 

\paragraph{Giant stars in IC~4651}
$T_{\rm eff}^{\rm LDR}$ is systematically lower than $T_{\rm eff}^{\rm SPEC}$ by an amount typically between 70 and 
90 K. The only exception is the star E95 for which the difference amounts to about 320 K. \cite{Pasqu04} already found 
for this star the largest difference between photometric and spectroscopic temperature among their giant star sample. 
However, the position on the HR diagram of this star corresponds to a subgiant and this could be the reason for the 
disagreement. 

\section{Color excess of IC~4651}
We can evaluate for each star the intrinsic color index $(B-V)_0$ by inverting for example the $(B-V)-T_{\rm eff}$ 
calibrations of \cite{Gray05} and \cite{Alonso96, Alonso99} with the aim to compute the color excess $E(B-V)$ of the 
cluster IC~4651. Thus, for the two temperature sets, $T_{\rm eff}^{\rm LDR}$ and $T_{\rm eff}^{\rm SPEC}$,
we obtain $E(B-V)\approx 0.12$ and 0.16 for the Gray's calibration, and $E(B-V)=0.13$ and 0.17 for the Alonso's 
calibration. It is worth noticing that there is not a large difference between color excesses 
obtained with the two calibrations. From a preliminary analysis, we find an improving of the agreement if we 
properly take into account the metallicity effects (\cite{Biazzo06c}). 
Moreover, our color excess values are in good agreement with the results of $E(B-V)=0.13$ and $E(b-y)=0.091$ 
obtained by \cite{Eggen71} and \cite{Pasqu04}, respectively ($E(b-y)=0.72E(B-V)$, \cite{Cardel89}). 
The present determination of $E(B-V)$ is a strong argument in favor of such a low reddening notwithstanding  
the distance of $\approx 900$\,pc estimated by \cite{Meibom02} and the low galactic latitude of IC~4651 
($\simeq 9^{\circ}$).

\section{Conclusion}
In this paper we have derived accurate atmospheric parameters for field evolved stars and giant stars in the open cluster 
IC~4651 by means of high-resolution spectra acquired with the ESO spectrographs FEROS and UVES. 

For the field giant stars, we find a good agreement between temperatures computed by \cite{daSil06}
with the curves-of-growth method and by ourselves with the LDR technique. For the giants in the 
intermediate-age open cluster IC~4651, we have determined the effective temperatures by means 
of the LDR method, that allowed us to compute the reddening. We find a rather low reddening towards 
the cluster, $E(B-V)\simeq 0.13$, that needs to be explained, given the high distance ($\simeq 900$\,pc) and
the low galactic latitude of IC~4651.

We conclude that our technique is well suited to derive accurate effective temperatures and reddening of clusters with a 
nearly-solar metallicity. The determination of very precise temperatures is of great importance to derive stellar age and 
mass distributions (\cite{Gira06}), representing a powerful tool for stellar population studies in addition to those 
based on photometric data. 

%
%

%
%

\end{document}